\documentclass[journal]{IEEEtran}
\ifCLASSINFOpdf
\else
\fi
\usepackage{graphicx}
\usepackage{times}
\usepackage{cite}
\usepackage{amssymb}
\usepackage{amsmath,epsfig}
\usepackage[font=footnotesize,labelfont=bf]{caption}
\usepackage{subcaption}
\usepackage{float}
\usepackage{siunitx}
\usepackage{setspace}

\usepackage{indentfirst}
\usepackage{algorithm}
\usepackage{algpseudocode}
\usepackage{bm}
\usepackage{graphicx,xcolor}
\usepackage{color}

\renewcommand{\vec}[1]{\mathbf{#1}}

\DeclareMathAlphabet{\mathpzc}{OT1}{pzc}{m}{it}
\usepackage{float}
\graphicspath{ {.} }
\begin{document}

\begin{table*}
\textcopyright{} 2020 IEEE.  Personal use of this material is permitted.  Permission from IEEE must be obtained for all other uses, in any current or future media, including reprinting/republishing this material for advertising or promotional purposes, creating new collective works, for resale or redistribution to servers or lists, or reuse of any copyrighted component of this work in other works.
\end{table*}
\thispagestyle{empty}
\pagebreak
\cleardoublepage
\setcounter{page}{1}

\title{Approximating the Hotelling Observer with Autoencoder-Learned Efficient Channels for Binary Signal Detection Tasks}

\author{Jason~L.~Granstedt,~
        Weimin~Zhou,~
        and~Mark~A.~Anastasio
\thanks{Jason L. Granstedt is with the Department of Computer Science, 
University of Illinois at Urbana-Champaign, Champaign, IL, 61820 USA
e-mail: (jasonlg@illinois.edu)}
\thanks{Weimin Zhou is with the Department of Electrical and Systems Engineering, 
Washington University in St.\ Louis, St.\ Louis, MO, 63130 USA
e-mail: (wzhou24@wustl.edu) }
\thanks{Mark A. Anastasio is with the Department of Bioengineering, 
University of Illinois at Urbana-Champaign, Champaign, IL, 61820 USA
e-mail: (maa@illinois.edu)}}


\maketitle

\begin{abstract}
The objective assessment of image quality (IQ) has been advocated for the analysis and optimization of medical imaging systems.
One method of obtaining such IQ metrics is through a mathematical observer.
The Bayesian ideal observer is optimal by definition for signal detection tasks, but is frequently both intractable and non-linear.
As an alternative, linear observers are sometimes used for task-based image quality assessment.
The optimal linear observer is the Hotelling observer (HO).
The computational cost of calculating the HO increases with image size, making a reduction in the dimensionality of the data desirable.
Channelized methods have become popular for this purpose, and many competing methods are available for computing efficient channels.
In this work, a novel method for learning channels using an autoencoder (AE) is presented.
AEs are a type of artificial neural network (ANN) that are frequently employed to learn concise representations of data to reduce dimensionality.
Modifying the traditional AE loss function to focus on task-relevant information permits the development of efficient AE-channels.
These AE-channels were trained and tested on a variety of signal shapes and backgrounds to evaluate their performance.
In the experiments, the AE-learned channels were competitive with and frequently outperformed other state-of-the-art methods for approximating the HO\@.
The performance gains were greatest for the datasets with a small number of training images and noisy estimates of the signal image.
Overall, AEs are demonstrated to be competitive with state-of-the-art methods for generating efficient channels for the HO and can have superior performance on small datasets.
\end{abstract}


%
\IEEEpeerreviewmaketitle

\section{Introduction}
Medical imaging systems are commonly optimized with consideration of a specific task~\cite{wagner1985unified}. Assessing performance of such systems requires an objective metric for image quality (IQ)~\cite{barrett2013foundations, kupinski2003ideal, park2007channelized, park2009efficient, shen2006using}.
For signal detection tasks, the Bayesian ideal observer (IO) has been advocated for producing a figure-of-merit for assessing IQ because it can maximize the amount of task-specific information in the measurement data~\cite{icru54report, barrett2013foundations,kupinski2003ideal, park2007channelized, park2009efficient,shen2006using}. 
For a binary signal detection task, the IO test statistic takes the form of a likelihood ratio.
Using this likelihood ratio as a test statistic in turn maximizes the area under the receiver operating characteristics (ROC) curve~\cite{barrett2013foundations, icru54report, kupinski2003ideal, park2007channelized}.
However, analytically determining the IO is generally difficult because it typically is a non-linear function and requires complete knowledge of the statistical properties of the image data.

There has been recent progress in developing approximations for computing the IO test statistic~\cite{park2009efficient, shen2006using}. One line of research involves sampling-based methods that utilize Markov-chain Monte Carlo methods to approximate the IO, but the work in this area has so far been limited to relatively simple object models~\cite{he2008toward, kupinski2003ideal, park2009efficient, abbey2008ideal}. Another recent development is the approximation of the IO with convolutional neural networks (CNNs)~\cite{zhou2019approximating}. An alternative method to approximating the IO's performance employs variational Bayesian inference~\cite{chen2019reconstruction}. This line of research has shown promise for implementing task-specific optimization of sparse reconstruction methods.

A common surrogate for the often-intractable IO is the Hotelling observer (HO)~\cite{barrett2015task,reiser2010task,sanchez2014task,glick2002investigation}.
The HO implements the optimal linear test discriminant for maximizing the signal-to-noise ratio of the test statistic~\cite{barrett1992linear, barrett1993model}.
Implementing the HO requires the estimation and inversion of a covariance matrix, which quickly grows as image size increases and can become intractable to compute~\cite{barrett2001megalopinakophobia}.
There are a few different strategies for mitigating the computational cost of inverting a large matrix~\cite{barrett2015task}.
One method is to avoid a direct inversion by implementing an iterative approach to estimate the test statistic~\cite{barrett2013foundations}.
If the measurement noise covariance matrix is known and an estimate of the background covariance matrix is available, covariance matrix decomposition is a viable option~\cite{barrett2013foundations} with the caveat that certain situations can lead to significant bias in the performance~\cite{kupinski2007bias}.
Alternatively, the test statistic can be learned directly from the images provided that there is a sufficient amount of data~\cite{zhou2019approximating, zhou2019learning}.
The most commonly employed method, however, is the implementation of channels that approximate the HO~\cite{barrett1998stabilized,gallas2003validating,park2009efficient}
These channels are linear transformations applied to reduce the dimensionality of the data, decreasing the computational costs of calculating the HO\@.

Conceptually, channels function by projecting the high-dimensional image data to a low-dimensional image manifold~\cite{tenenbaum1998mapping}.
Image data frequently can be compressed into a reduced-dimensionality manifold~\cite{beymer1996image, turk1991eigenfaces}.
Ideally, the manifold embedding would preserve the important features of the data~\cite{weinberger2006unsupervised}.
This projection operation is defined by the channel matrix.
Channels are known as efficient if they approximate the original observer's performance while reducing the dimensionality of the data~\cite{myers1987efficient}.
Prior work in computing efficient channels includes Laguerre-Gauss (LG)~\cite{gallas2003validating}, singular value decomposition (SVD)~\cite{park2009singular}, partial least squares (PLS)~\cite{witten2010pls}, and the filtered channel observer (FCO)~\cite{diaz2015derivation}.
In addition to learning efficient channels, there are approaches that seek to mimic the human observer's performance.
An early approach learned the relationship between channel features and human observer performance with a support vector machine~\cite{jorvan2009human}.
Another approach investigated optimization with respect to the HO and human observers on accelerated MRI reconstruction~\cite{angel2019lg}.

Autoencoders (AEs) are a type of artificial neural network (ANN) that are characterized by a mirror structure, with the target output of the network similar to the input~\cite{rumelhart1986learning, hinton2006fast, hinton2006reducing, bengio2007scaling, erhan2010unsupervised}.
They are designed to learn a lower-dimensional representation of the data called an embedding.
The portion of the network that transforms the input to the embedding is known as the encoder, and the portion that transforms the embedding back into the original data space is known as the decoder.
A good embedding is capable of significant data compression while retaining most of the information from the original data.
The data compression qualities of AEs make them desirable to use in many tasks, and they have been applied in state-of-the-art systems for classification~\cite{vincent2008extracting}, noise reduction~\cite{nishio2017conv}, and regression~\cite{zhang2017age}.
The widespread success of the AE is due to its ability to generate low-dimensional representations of images, which increases the efficiency of further processing by attenuating noise and embedding data to its most important components.
In general, a linear AE with optimal weights projects the data onto a subspace spanned by its top principle directions~\cite{kunin2019pca}.

In this work, the problem of learning task-informed embeddings with an AE is explored.
The AE is modified to learn the optimum transformation matrix that maximizes the amount of task-specific information encoded in its latent states.
This learning task is demonstrated to be equivalent to learning efficient channels for the HO\@.
To the best of our knowledge, this is the first time a connection has been established between numerical observer channels and autoencoder-generated embeddings.
The considered model is a linear AE with one hidden layer and one set of tied weights, as described below.
Numerical studies are performed with binary signal detection tasks that involve a range of signals and backgrounds.
The performance of the AE-learned channels in these studies is compared to state-of-the-art channelized methods.
The potential advantages and limitations of this new approach are also discussed.

The remainder of this work is organized as follows. 
In Sec.~\ref{sec:background} an overview of binary signal detection theory is presented. The HO, CHO and AE are also reviewed in that section.
A novel methodology for learning channels for the location-known binary signal detection task using an AE is developed in Sec.~\ref{sec:AETSI}.
The numerical studies and results of the proposed method for approximating the HO are included in Secs.~\ref{sec:num} and~\ref{sec:results}, along with a comparison to other state-of-the-art methods.
Finally, the paper concludes with a discussion of the work in Sec.~\ref{sec:concludes}.
 
\section{Background}\label{sec:background}
Consider the linear digital imaging system
\begin{equation} \label{eq:dim}
\vec{g} = \mathcal{H}{f}(\vec{r}) + \vec{n}, 
\end{equation}
where \(\vec{g}\in\mathbb{R}^{n}\) is the measured image data vector,
\(\mathcal{H}\) denotes a continuous-to-discrete (C-D) imaging operator that maps \(\mathbb{L}_2(\mathbb{R}^2)\rightarrow\mathbb{R}^{n}\), 
 \({f}(\vec{r})\) is the object function with the 2-D spatial coordinate \(\vec{r}\),
 and \(\vec{n}\in\mathbb{R}^{N}\) is the random measurement noise.
The object function \(f(\vec{r})\) will be abbreviated as \(\vec f\) and can be either deterministic or stochastic,
depending on the specification of the signal detection task.

\subsection{Formulation of binary signal detection tasks}
The binary signal detection task considered involves the classification of an image by an observer into one of two hypotheses: signal-present (\(H_1\)) or signal-absent (\(H_0\)). 
The imaging processes under these two hypotheses can be described as

\begin{subequations}\label{eq:imgH}
\begin{equation}
H_{0}: \vec{g} = \mathcal{H}\vec{f}_{b} + \vec{n},
\end{equation}
\begin{equation}
H_{1}:  \vec{g} = \mathcal{H}(\vec{f}_{b}+\vec{f}_{s}) + \vec{n},
\end{equation}
\end{subequations}
where \(\vec{f}_{b}\) and \(\vec{f}_{s}\) represent a background and a signal object, respectively. Depending on the imaging task, these can be either random or fixed.

To perform a binary signal detection task, an observer computes a test
statistic \(t(\vec{g})\) that maps the measured image \(\vec{g}\) to a real-valued scalar variable.
This scalar is compared against a threshold \(\tau\) to classify \(\vec{g}\) as satisfying either \(H_0\) or \(H_1\).
To determine the desired performance for the signal detection task, a receiver operating characteristic (ROC) curve can be plotted to depict the trade-off between the false-positive fraction (FPF) and the true-positive fraction (TPF) by varying the threshold \(\tau\). The overall signal detection performance of the observer can be summarized by computing the area under the ROC curve (AUC)~\cite{metz1986roc}.

\subsection{Bayesian Ideal Observer and Hotelling Observer}\label{subsec:IO}
The IO is optimal and sets the upper limit for observer performance on binary signal detection tasks.
The IO test statistic is defined as any monotonic transformation of the likelihood ratio \(\Lambda_\text{LR}(\vec{g})\) that takes the form~\cite{barrett2013foundations, kupinski2003ideal, kupinski2001ideal}
\begin{equation}
\Lambda_\text{LR}(\vec{g}) = \frac{p(\vec{g}|H_1)}{p(\vec{g}|H_0)},
\end{equation}
where \(p(\vec{g}|H_0)\) and \(p(\vec{g}|H_1)\) are conditional probability density functions
 that describe the measured data \(\vec{g}\) under hypothesis \(H_0\) and \(H_1\), respectively.

An alternative to the IO for assessing signal detection performance is the HO\@. The HO test statistic is defined as 
\begin{equation}
t(\vec{g}) =  \vec{w}_{HO}^T \vec{g},
\end{equation}
where \(\vec{w}_{HO}\in\mathbb{R}^{M}\) is the observer template.
Let \(\bar{\vec{g}}(\vec{f})\equiv\langle \vec{g} \rangle_{\vec{g}|\vec{f}}\)
denote the conditional mean of the image data given an object function.
Similarly, let \(\bar{\vec{g} }_j\equiv \langle\langle \vec{g} \rangle_{\vec{g}|\vec{f}} \rangle_{\vec{f}|H_j}\)
denote the conditional mean averaged with respect to object randomness associated with \(H_j\).
The Hotelling template \(\vec{w}_\text{HO}\) is defined as\cite{barrett2013foundations}
\begin{equation} \label{eq:w_HO}
\vec{w}_\text{HO}=\left[\frac{1}{2}(\vec{K}_0 + \vec{K}_1)\right]^{-1}\Delta{\bar{\vec{g}}},
\end{equation}
where
\begin{equation}
\vec{K}_j =\left\langle \langle [\vec{g} -{\bar{\vec{g} }}_j][\vec{g} -{\bar{\vec{g} }}_j]^{T}  \rangle_{\vec{g}|\vec{f}} \right\rangle_{\vec{f}|H_j},
\end{equation}
\begin{equation}
\Delta {\bar{\vec{g}}} = {\bar{\vec{g} }}_1 - {\bar{\vec{g} }}_0.
\label{eq:sigest}
\end{equation}
Here, \(\vec{K}_j\) is the covariance matrix of the measured data \(\vec{g}\) under the hypothesis \(H_j\) and \(\Delta{\bar{\vec{g}}}\) is the difference between the mean of the measured data \(\vec{g}\) under the two hypotheses. 

The signal to noise ratio (SNR) associated with the test statistic \(t\) is another commonly employed FOM for assessing signal detection performance and is given by~\cite{barrett1993model}
\begin{equation}\label{eq:SNR2}
\text{SNR}_t = \frac{\langle t\rangle_1 - \langle t\rangle_0}{\sqrt{\frac{1}{2}\sigma_0^2+\frac{1}{2}\sigma_1^2}},
\end{equation}
where \(\langle t \rangle_j\) and \(\sigma_j^2 = \left\langle(t-\langle t\rangle_j)^2 \right\rangle_j\) are the mean and variance of \(t\) under the hypothesis \(H_j\) (\(j=0, 1\)). While the IO maximizes the AUC of an observer, the HO maximizes the SNR of the test statistic \(t(\vec{g})\)~\cite{barrett1993model}.

\subsection{Channels}
Computation of the HO can become intractable for large image sizes due to the cost of inverting the covariance matrix in Eqn. (\ref{eq:w_HO}). Additionally, it may be difficult to estimate a full rank covariance matrix in limited-data cases. To mitigate this problem, a channelized version of the image \(\vec{g}\) can be introduced as~\cite{barrett1993model}
\begin{equation}
\vec{v} = \vec{Tg} = \vec{T}\left(\mathcal{H}\vec{f} + \vec{n}\right),
\label{eq:CHO}
\end{equation}
where \(\vec{v}\) is a \(M\times 1\) channel-reduced image and \(\vec{T}\) is a \(M\times N\) matrix.
The number of channels, \(m\), determines the dimensionality reduction from the original data of size \(n\)
Applying the HO to the channel-reduced data yields the channelized HO (CHO)~\cite{barrett1993model}, with the test statistic taking the form of 
\begin{equation}
\label{eq:HOv}
T(\mathbf{v}) = \vec{w}_v^T\vec{v}=\left( \mathbf{K}^{-1}_\mathbf{v} \Delta \mathbf{\bar{v}} \right)^T \mathbf{v}.
\end{equation}
Here, \(\vec{K}_\vec{v} = \frac{1}{2}\left[ \vec{K}_{\vec{v},1} + \vec{K}_{\vec{v},0} \right]\) and \(\Delta \mathbf{\bar{v}} = \mathbf{\bar{v}}_1 - \mathbf{\bar{v}}_0\), where \(\vec{K}_{\vec{v},j} = \langle\langle\left[\vec{v} - \vec{\bar{v}}_j \right]\left[\vec{v} - \vec{\bar{v}}_j \right]^T\rangle_{\vec{v}|\vec{f}}\rangle_{\vec{f}|H_j}\) and \(\mathbf{\bar{v}}_j = \langle\langle\vec{v}\rangle_{\vec{v}|\vec{f}}\rangle_{\vec{f}|H_j}\) for \(j = (0,1)\). 

It is desirable to minimize the number of channels to maximize computational efficiency, since the dimensionality of \(\vec{K}_\vec{v}\) is proportional to the number of channels.
However, these channels should maximize the retained, task-relevant, information to provide an efficient approximation of the HO\@. There are several methods that exist for selecting efficient channels. One of the first was LG channels~\cite{myers1987efficient}. These channels are a combination of a Gaussian function with a Laguerre polynomial and were proposed due to their structural similarity with the Hotelling template for certain detection tasks. These channels are suitable for a smooth rotationally symmetric signal on a lumpy background, but may have suboptimal performance for arbitrary signals and more complex backgrounds~\cite{witten2010pls}.

An alternative to LG channels are SVD channels~\cite{park2009singular}. These channels are singular vectors that form a basis for image vectors in the range of the imaging operator. The most efficient set of channels constructed from this method involved decomposing the noiseless signal image by use of the singular vectors  and choosing the top \(m\) of them to form the channel set. However, this method is computationally expensive and system-specific.

Two current state-of-the-art methods for generating efficient channels that work on arbitrary signals and backgrounds without any specific knowledge of the imaging system are partial least squares (PLS)~\cite{witten2010pls} and filtered channel observer (FCO)~\cite{diaz2015derivation}. PLS applies a data reduction technique that iteratively constructs a number of latent vectors that maximize the covariance between the data and the true image labels. PLS represents an attractive method to use in limited-data cases and/or large image sizes and works well with noisy and heavily correlated data. However, the technique suffers a notable degradation of performance when the amount of available image data is small~\cite{witten2010pls}.

FCO channels were initially developed as anthropomorphic channels to approximate human signal detection performance for irregularly-shaped signals~\cite{diaz2015derivation}. However, FCO channels have been explored as efficient channels for the HO\cite{diaz2015derivation,badano2018victre}. The FCO convolves a selected set of baseline channels with the signal before computing the observer template. For this work, LG channels were selected as the baseline set of channels due to both LG's past success~\cite{myers1987efficient} and similar decisions with the FCO method in more recent work~\cite{badano2018victre}. This realization of the FCO method will be referred to as convolutional LG\@.

\subsection{Neural Networks for Approximating the IO} A feed-forward ANN is a system of computational units associated with tunable parameters called weights~\cite{schmidhuber2015deep, lecun2015deep}. 
A feed-forward ANN is capable of approximating any continuous function if it has a sufficiently complex architecture~\cite{hornik1989multilayer, hornik1991universal}.
ANNs have been employed to form numerical observers, with the focus on directly estimating the test statistic~\cite{kupinski2001ideal, zhou2018learning, zhou2019approximating}.
Kupinski~\emph{et al.}~\cite{kupinski2001ideal} utilized conventional fully connected neural networks to approximate the IO on low-dimensional extracted image features.
Zhou and Anastasio extended this work to higher-dimensional data and allowed for native processing of image data by replacing the FCNN with a convolutional neural network~\cite{zhou2018learning, zhou2019approximating}.
However, both of these approaches focus on learning the test statistic directly and may require a large amount of training data to accurately approximate the IO\@.

\subsection{Autoencoders}
A specialized type of ANN is the autoencoder (AE)~\cite{rumelhart1986learning, hinton2006fast, hinton2006reducing, bengio2007scaling, erhan2010unsupervised}.
The AE is characterized by a mirror structure, with the input of the network similar to the target output.
An AE has three distinct components: an encoder, an embedding, and a decoder.
The encoder transforms the input to the embedding, which generally has a significantly reduced dimensionality compared to the input.
The decoder transforms the embedding into the target output.
In a canonical AE, the decoder is specified to reconstruct an approximation of the input to the encoder.
AEs are frequently employed for their data compression properties in state-of-the-art systems for classification~\cite{vincent2008extracting}, regression~\cite{zhang2017age}, noise reduction~\cite{nishio2017conv}, anomaly detection~\cite{chandola2009anomaly}, and image recovery~\cite{mousavi2015recovery} tasks.
Additional performance improvements can be made by injecting additional information into the AE training process.
Studies have shown that exploiting \textit{a priori} information through implicitly defined nonparametric functions can introduce task-specific information in the training of AEs~\cite{snoek2012nonparametric1, snoek2012nonparametric2}.

In contrast to previous work with ANNs, an AE is usually trained in an unsupervised way~\cite{vincent2008extracting}.
One aspect of AEs that has recently been considered is the concept of tied weights~\cite{vincent2010stacked}. Tied weights further enforce the mirror-like structure of the AE by forcing the encoder and decoder matrices to be symmetric. Tied-weight AEs have been shown to perform similarly to untied-weight AEs, but require less data to train because of the reduction in parameters.

In general, the layers in an AE specify many sets of matrix multiplications with added bias terms and nonlinear transformations.
By restricting the operations to only matrix multiplications, linear AEs can be obtained.
In these cases, the encoder and decoder can each be described by a transformation matrix that transforms to or from the data embedding.
Such a simplified network is considered in this work since this configuration's encoder has a natural parallel with the channel matrix in the CHO\@.
The input to the network is a noisy image and the target output is either the input image or a related version of the input image, depending on the task.

An optimization problem is solved to determine the weights of the AE by minimizing a reconstruction loss.
The solution of the optimization problem is computed by minimizing the loss function using a variation of the backpropogation algorithm~\cite{kingma2014adam}.
The traditional loss function for an AE is the mean squared error between the input and the output of the network. Given \(N\) vectorized background images \(\vec{g}_i\) of size \(n^2\times 1\), the traditional loss function corresponding to a zero-bias linear AE is~\cite{rumelhart1986learning}
\begin{equation}
L_{trad} (\mathbf{W}_1, \mathbf{W}_2) = \frac{1}{N}\sum_{i=1}^N \left\Vert\mathbf{W}_2\mathbf{W}^T_1\mathbf{g}_i - \hat{\mathbf{g}}_i\right\Vert_2^2,
\label{eq:trad_loss}
\end{equation}
where \(\mathbf{W_1}\) and \(\mathbf{W_2}\) are each \(n\times m\) weight matrices that parameterize the encoder and decoder of the AE, respectively.
The target reconstruction is represented by \(\hat{\mathbf{g}}_i\), which can be the same as or different from the input data but is usually closely related.
For example, in denoising problems the target output is a clean version of the input image.


\section{Method - Autoencoder-Learned Channels}\label{sec:AETSI}
\begin{figure*}
\centering
\includegraphics[width=1.0\linewidth]{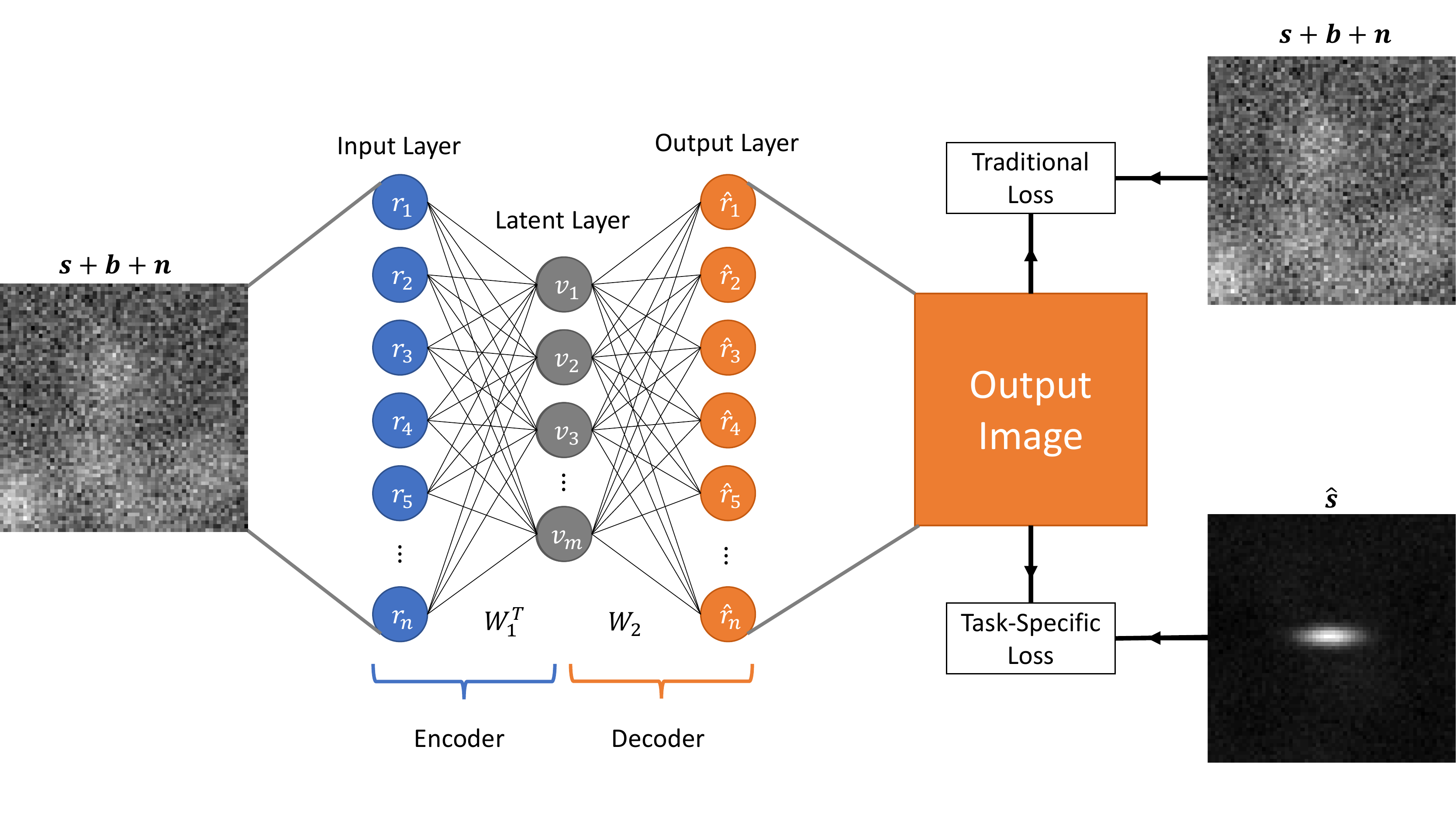}
\vspace{-1.1cm}
\caption{Diagram of the proposed method. A noisy image is the input to an encoder, which is mapped to an \(m\)-dimensional latent space to encode the information. The encoder transformation matrix is given by \(\vec{W}_1^T\). The embedded representation is then multiplied by the decoder transformation matrix, \(\vec{W}_2\), to return to image space and generate the output image. Two different loss functions are considered for training this model. The traditional loss function computes the MSE between the output image and the input image. This approach attempts to reconstruct the entire input image. The second considered loss function is the task-specific loss, which calculates the difference between the output image and estimated signal image \(\hat{\vec{s}}\). This loss maximizes the signal-specific information of the input image.}
\label{fig:AE_Graphic}
\end{figure*}

\begin{figure}
\centering
\begin{subfigure}{0.11\textwidth}
\includegraphics[width=1.0\linewidth]{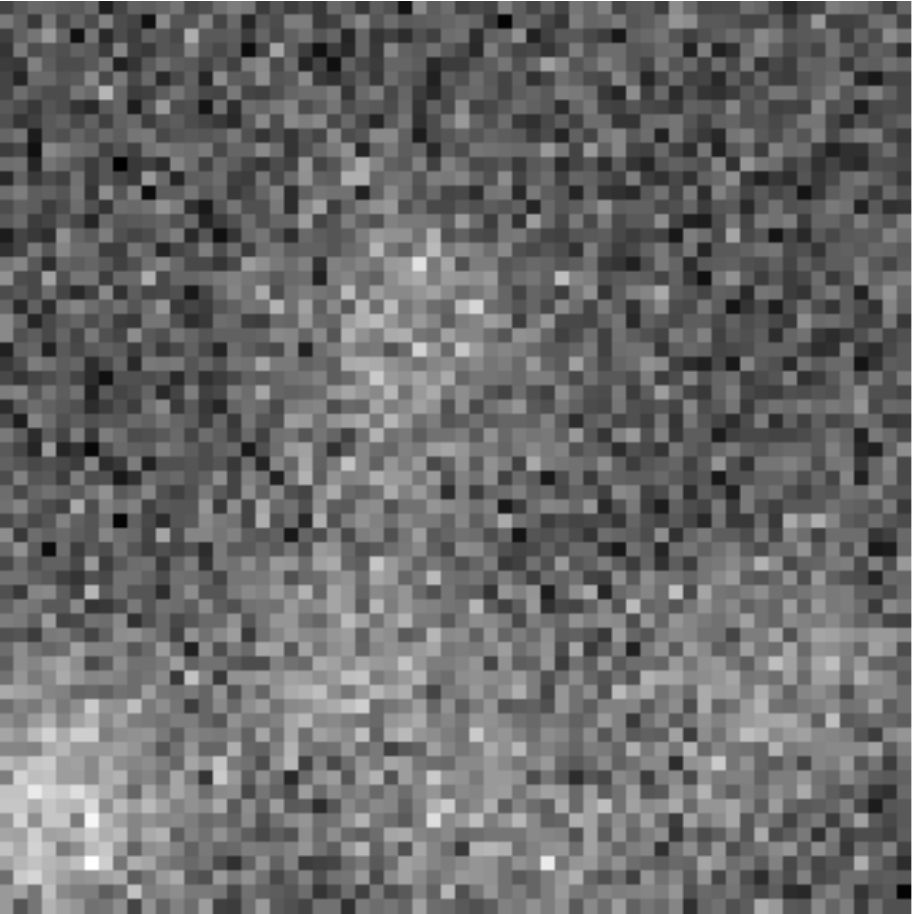}
\caption{}
\end{subfigure}
\begin{subfigure}{0.11\textwidth}
\includegraphics[width=1.0\linewidth]{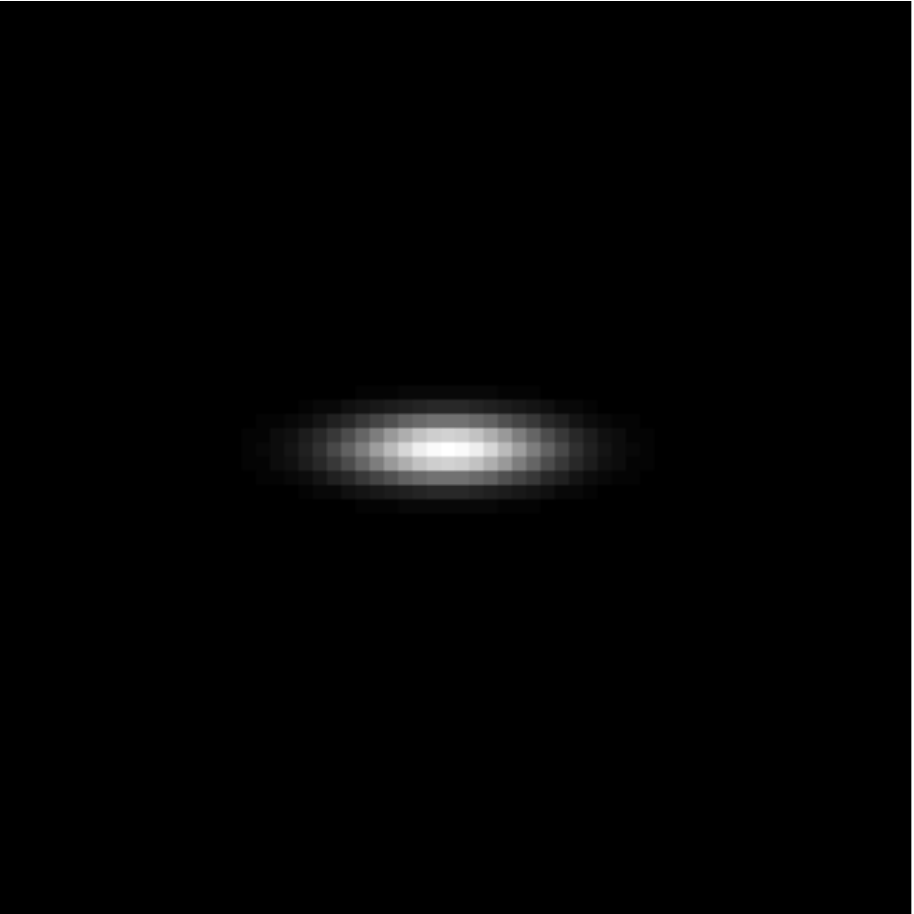}
\caption{}
\end{subfigure}
\begin{subfigure}{0.11\textwidth}
\includegraphics[width=1.0\linewidth]{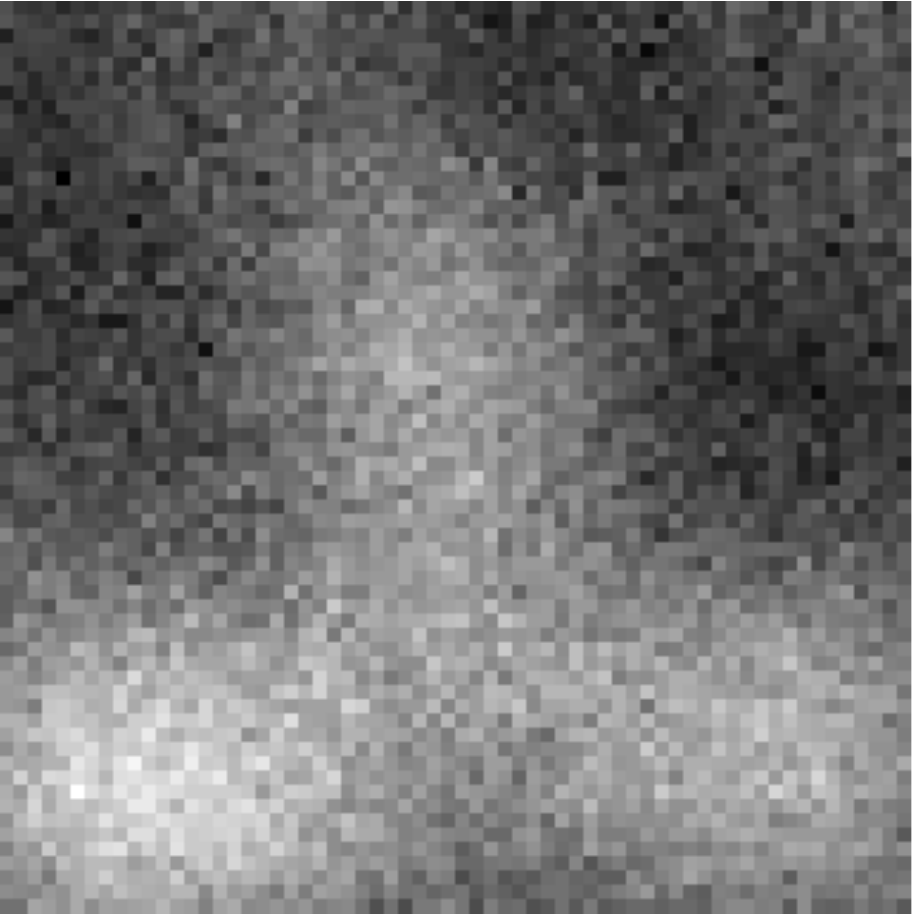}
\caption{}
\end{subfigure}
\begin{subfigure}{0.11\textwidth}
\includegraphics[width=1.0\linewidth]{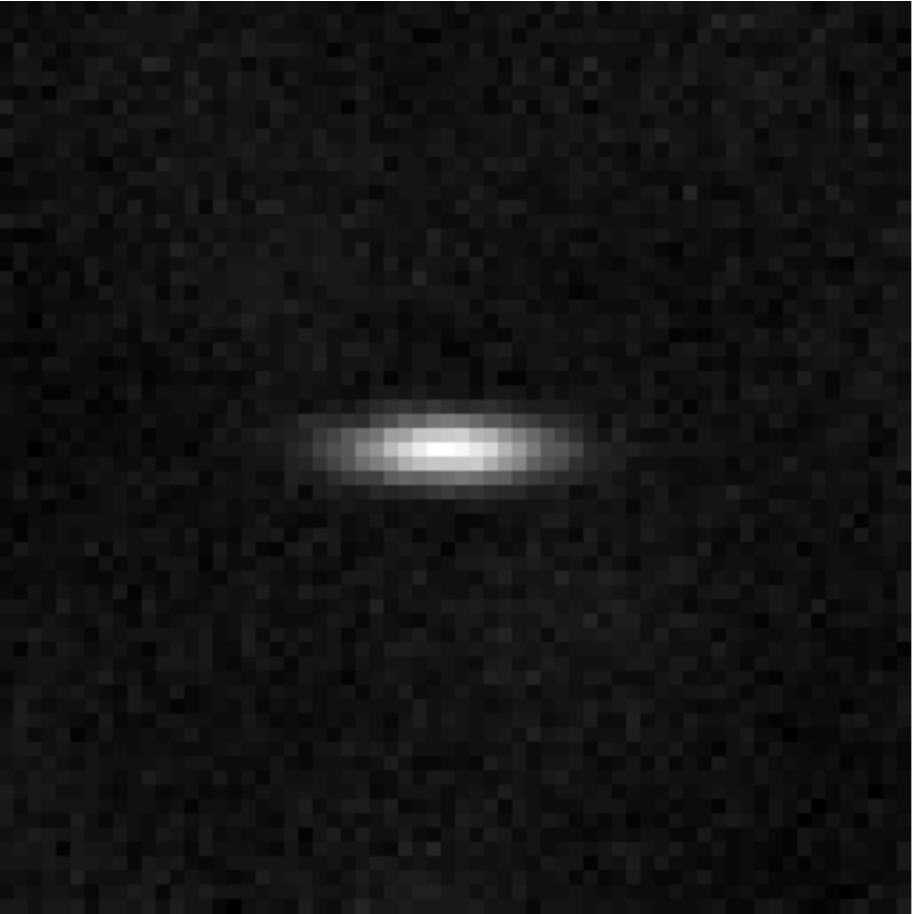}
\caption{}
\end{subfigure}
\caption{Reconstructed images corresponding to each of the loss functions. The grayscale in each case is adjusted to maximize visibility. (a) is the input image, which contains the faint Gaussian elliptical signal in (b). The traditional AE with 20 channels reconstructs the image in (c) while the task-specific tied-weight AE with 10 channels reconstructs the image in (d). Note that the reconstructed image in (c) is noticeably less noisy than the input image (a), which it is attempting to reconstruct. This is due to the limited number of latent states in the model embedding the largest structures in the input images. Noise cannot be effectively encoded for an image, so it is attenuated.}
\label{fig:samp_recon}
\end{figure}

A method for learning efficient channels for the CHO with an AE is described below.
A connection between AE weights and the CHO framework is established to illustrate the connection between the learned data embeddings and more traditional channels.

\subsection{Autoencoder Channels and Linear Autoencoders}
The learned weights of an AE have an additional interpretation when considered in the framework of a signal detection task.
The weights define a mapping from the high-dimensional image space to a low-dimensional embedding space.
This is conceptually equivalent to the CHO channel matrix \(\vec{T}\).
The AE weights can be employed as channels for the CHO by setting \(\vec{T}=\vec{W}_1^T\) in Eqn. (\ref{eq:CHO}).
Intuitively, these AE-learned channels capture the data most important for reconstructing the image.
%
%
%

The loss function in Eqn. (\ref{eq:trad_loss}) causes the AE to encode the entirety of the input image.
This makes the traditional AE suboptimal for learning channels because a significant portion of the data embedding is dedicated to reconstructing certain components of the background and noise that may not be highly relevant to the detection task.
To circumvent this, as described below, information about the signal can be incorporated into the AE training process to preserve task-specific information.

\subsection{Task-Specific Autoencoders}
A novel modification to the loss function to improve the learned data embedding and resulting signal detection performance for AE-channels is presented here.
Ideally, the entirety of the AE embedding would be dedicated to the task-specific information.
This would minimize the proportion of the embedding that is dedicated to extraneous information and lead to a more efficient set of channels.
By changing the AE's target reconstruction to just the mean signal image, the background and noise are suppressed during the reconstruction process.
This results in an embedding in which the signal can be accurately represented.
This new approach minimizes the MSE between the reconstructed image and the estimated signal image and takes the form of
\begin{equation} \label{eq:AEmod}
L_{task}(\mathbf{W}_1, \mathbf{W}_2) = \frac{1}{N}\sum_{i=1}^N \left\Vert\mathbf{W}_2\mathbf{W}^T_1\mathbf{g}_i - I \left( \mathbf{g}_i \right) \Delta \mathbf{\bar{g}}\right\Vert_2^2,
\end{equation}
where \(\Delta \mathbf{\bar{g}}\) is defined in Eqn. (\ref{eq:sigest}) and \(I(\cdot)\) is the indicator function that returns 1 if the signal is present and 0 otherwise.
Note that this loss function uses label information, and thus is a supervised learning algorithm.
Considering the background as noise permits the entire capacity of the embedding to focus on the task-specific information.
Using the signal template as the target image assists the training process in identifying an embedding that preserves task-specific information.
The indicator function and alteration to the desired output also breaks the traditional AE's connection to principle directions~\cite{kunin2019pca}.
As shown below, this modification to the loss function is capable of generating efficient channels for the CHO\@.
A diagram of the AE with both the traditional and task-based approach for the signal detection task is provided in Fig.~\ref{fig:AE_Graphic}, with a sample reconstruction from AEs trained using both loss functions shown in Fig.~\ref{fig:samp_recon}.
Both the task-specific and traditional loss functions can be minimized by use of a gradient-descent method, with specific implementation details provided in Sec.~\ref{expparam}.

 \section{Numerical studies}\label{sec:num}
Numerical simulation studies were conducted to evaluate the performance of the proposed method for learning efficient channels for the CHO\@. 
All simulations addressed background-known-statistically (BKS) signal detection tasks.
Four distinct binary signal detection tasks were considered.
Using a lumpy background, a location-known task and signal-known-statistically task were considered.
These tasks enabled the HO to be determined both using covariance matrix decomposition~\cite{barrett2013foundations} and direct computation according to Eqn.~(\ref{eq:w_HO}).
These observers will be referred to as HO-CMD and HO-Direct, respectively.
On a breast phantom background, two location-known signal detection tasks using signals of different shapes and sizes were considered.
These tasks allowed for the evaluation of channelized methods on a more realistic medical imaging task.
ROC curves were fit by use of a binormal model~\cite{metz1986roc, metz1999proper, pan1997proper} with the fitted AUC values reported.
The experimental results are reported in distinct sections based on the image background model, with the details for each signal detection task and the training of neural networks are given in the appropriate subsections.

\subsection{Signal detection tasks that utilize a lumpy background model}\label{sec:exp1}
Two different signal detection tasks were performed on a lumpy background model~\cite{rolland1992effect} with an idealized parallel-hole collimator system~\cite{kupinski2003ideal}. 
Further details about each of the components is provided below.

\subsubsection{Lumpy Background}
A stochastic lumpy object model was used as the background~\cite{rolland1992effect}
\begin{figure}
\centering
\begin{subfigure}{0.11\textwidth}
\includegraphics[width=1.0\linewidth]{s-eps-converted-to.pdf}
\caption{}
\end{subfigure}
\begin{subfigure}{0.11\textwidth}
\includegraphics[width=1.0\linewidth]{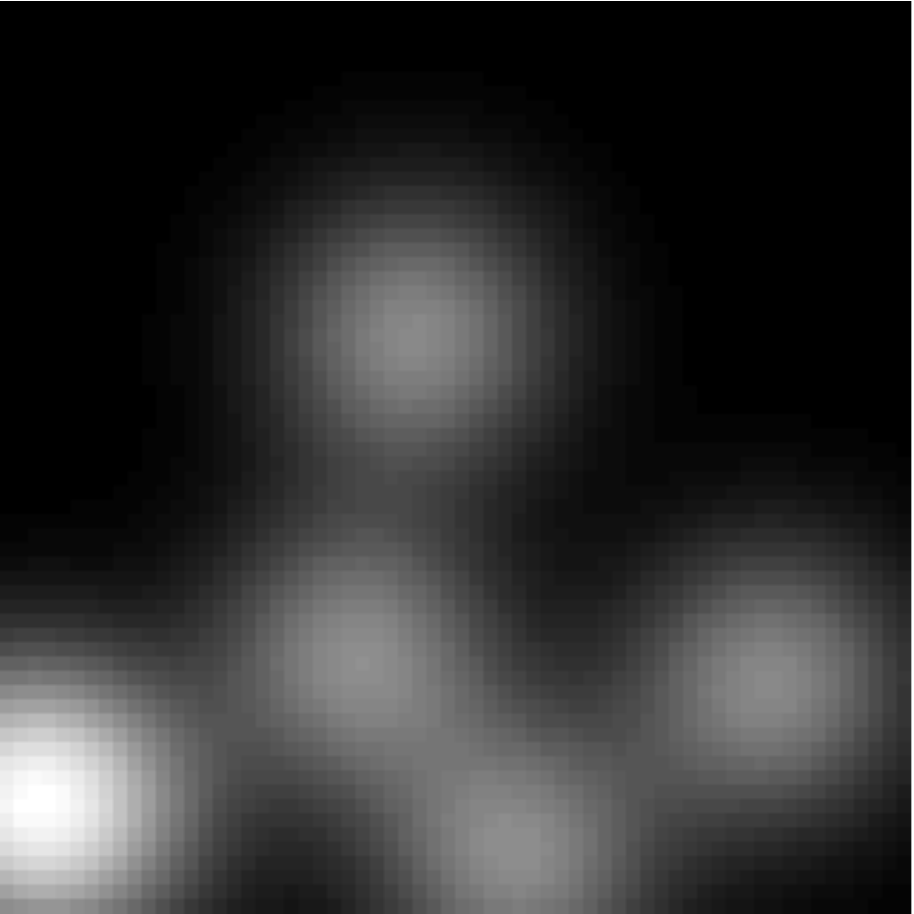}
\caption{}
\end{subfigure}
\begin{subfigure}{0.11\textwidth}
\includegraphics[width=1.0\linewidth]{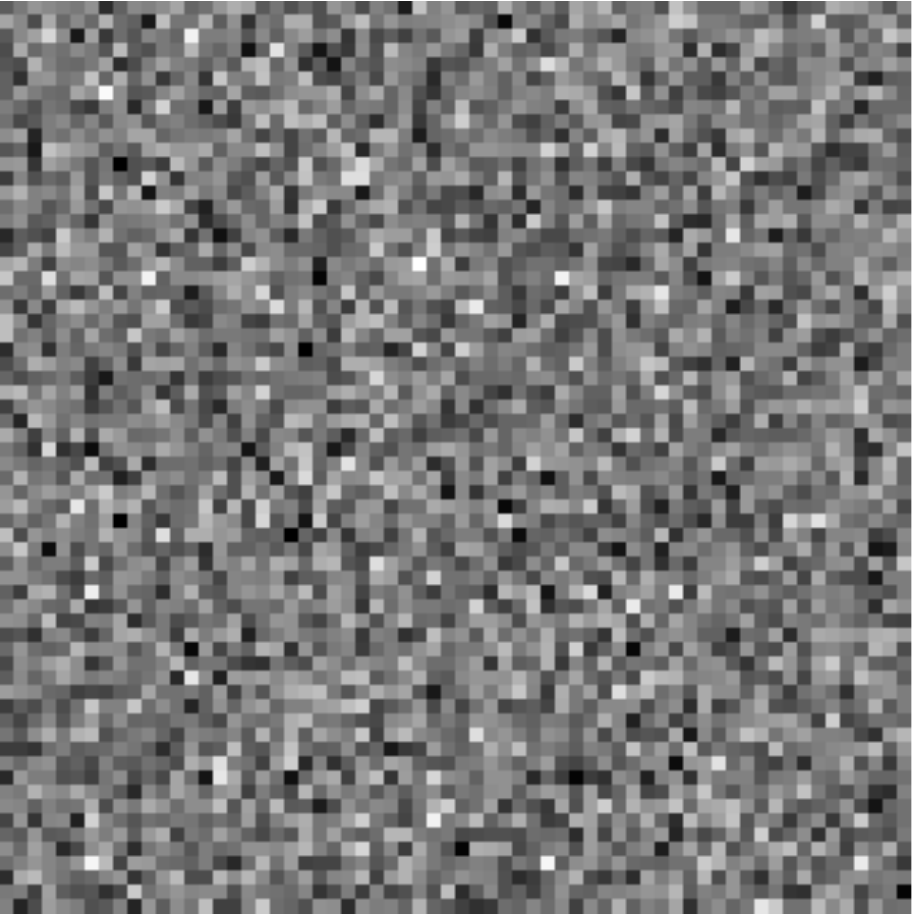}
\caption{}
\end{subfigure}
\begin{subfigure}{0.11\textwidth}
\includegraphics[width=1.0\linewidth]{sbn-eps-converted-to.pdf}
\caption{}
\end{subfigure}
\caption{Sample generation for signal-present images used in the lumpy background experiments. The grayscale in each case was adjusted to maximize visibility. The signal image (a) was added to the lumpy background (b) and the Gaussian noise (c) to produce the composite dataset image (d). The signal image is an elliptical Gaussian signal with width \(\sigma_x=5, \sigma_y=1.5\).}
\label{fig:circ_image_gen}
\end{figure}
 \begin{equation} \label{eq:lumpy}
\vec{f}_{b}(\vec{r}) = \sum_{n=1}^{L}l(\vec{r}-\vec{r}_{n} | a, s),
\end{equation}
where \(L\sim Poiss\left(\overline{L}=5\right)\) is the number of lumps that is sampled from Poisson distribution with the mean set to 5
and \(l(\vec{r}-\vec{r}_n | a, s)\) is the lumpy function modeled by a symmetric 2D Gaussian function with amplitude \(a\) and width \(s\)
\begin{equation}
l(\vec{r}-\vec{r}_n | a, s) = a\exp{\left(-\frac{(\vec{r}-\vec{r}_n)^T(\vec{r}-\vec{r}_n)}{2s^2} \right)}. 
\end{equation}
Here, \(\vec{r}_{n}\) is the uniformly-sampled position of the \(n^{th}\) lump. The magnitude and width of the lumps were set to the frequently-employed values of \(a=1\) and \(s=7\).
An example of a signal-present image in the dataset with a circular signal is located in Fig.~\ref{fig:circ_image_gen}.

\subsubsection{Imaging system}\label{sec:signal}
The stylized imaging system in these studies was a linear C-D mapping describing an idealized parallel-hole collimator system with a point response function\cite{kupinski2003ideal, kupinski2003experimental}
\begin{equation} \label{eq:kernel}
\vec{h}_m(\vec{r}) = \frac{h}{2\pi w^2} \exp{\left(-\frac{\left(\vec{r}-\vec{r}_m\right)^{T}(\vec{r}-\vec{r}_m)}{2w^2}\right)},  
\end{equation}
with the height \(h = 40\) and the width \(w=0.5\).

\subsubsection{Signals}
The signal function \(\vec{f}_s(x, y)\) was a 2D Gaussian function
 \begin{equation}\label{eq:ske_s}
 \vec{f}_s(\vec{r}) = A  \exp{\left(-\left(\vec{R_\theta}\left(\vec{r} - \vec{r_c}\right)\right)^{T} \vec{D^{-1}} \left(\vec{R_\theta}\left(\vec{r} - \vec{r_c}\right)\right)\right)},
 \end{equation}
 where \(A=0.2\) is the amplitude and \(\vec{r_c}\) is the coordinate of the signal location.
 Here, \(R_\theta\) is the Euclidean rotation matrix that rotates the Gaussian by an angle of \(\theta\) and is given by
 \begin{equation}
  \vec{R_\theta} = \begin{bmatrix}
    \cos\theta & -\sin\theta \\
    \sin\theta & \cos\theta
  \end{bmatrix},
 \end{equation}
 and \(\vec{D}\) is a scaling matrix that controls the width of the Gaussian along each axis and is given by
 \begin{equation}
  \vec{D} = \begin{bmatrix}
    2\sigma_x^2 & 0 \\
    0 & 2\sigma_y^2
  \end{bmatrix}.
 \end{equation}
For both experiments involving the lumpy background, the elliptical Gaussian signal was set to have the parameters \(\sigma_x=5\) and \(\sigma_y=1.5\).
The image size was selected to be \(64\times 64\) with the signal centered at \(\vec{r_c} = \left[32, 32\right]^T\).
The value of \(\theta\) varied depending on the type of task.
%
 %
\subsubsection{Detection Tasks}
The first signal detection task employed \(\theta = 0\), forcing the signal to take the same orientation in each image. Thus, the signal location and shape were fixed. The signal template was computed according to Eqn. (\ref{eq:sigest}), which resulted in a noisy estimate of the signal. 
The second signal detection task sampled \(\theta\) uniformly from the set \([\ang{0}, \ang{45}, \ang{90}, \ang{135}]\). This allowed for four distinct orientations of the elliptical Gaussian. The mean signal was also computed with Eqn. (\ref{eq:sigest}), which resulted in a noisy estimate of the signal averaged across the four possible realizations.
\subsubsection{Dataset Generation}
A training set of 60\thinspace000 unique background images with noise were generated for the lumpy object model.
The background images were generated separately from the signal image in Eqn. (\ref{eq:ske_s}) using the appropriate background model.
Each background image was summed with a unique noise vector drawn from an i.i.d. Gaussian distribution \({n}_m \sim \mathcal{N}\left(0, \delta^2\right)\) with a mean of 0 and standard deviation \(\delta=20\).
These images were then paired, with half designated for signal present and half for signal absent.
Each signal present image was summed with the signal image to generate the final training data set of 30\thinspace000 paired images.
Another set of 5000 paired images was generated for determining the channel covariance matrix after the channels had been learned and a further set of 5000 paired images were held out as a testing dataset.
 
\subsection{Location-known tasks that utilize a breast phantom dataset}\label{sec:exp2}
Two further signal detection tasks were performed on a breast phantom background employing the VICTRE dataset~\cite{badano2018victre}.
This dataset contains simulated digital mammography (DM) images and was employed previously in a location-known human observer study to evaluate imaging systems~\cite{badano2018victre}.
The images are divided into four categories of breast types of decreasing difficulty for lesion detection: extremely dense, heterogeneously dense, scattered fibroglandular, and fatty.
The signals in the dataset are microcalcification clusters and spiculated masses.
For each signal, there are associated signal-absent and signal-present images.
The signal remains constant in location and shape throughout all the signal-present images, but a clean signal image is not available. 
An estimation of the signal is obtained from the difference of the mean signal-present and signal-absent images according to Eqn. (\ref{eq:sigest}), making this a location-known task~\cite{badano2018victre}.

\begin{figure}
\centering
\begin{subfigure}{0.11\textwidth}
\includegraphics[width=1.0\linewidth]{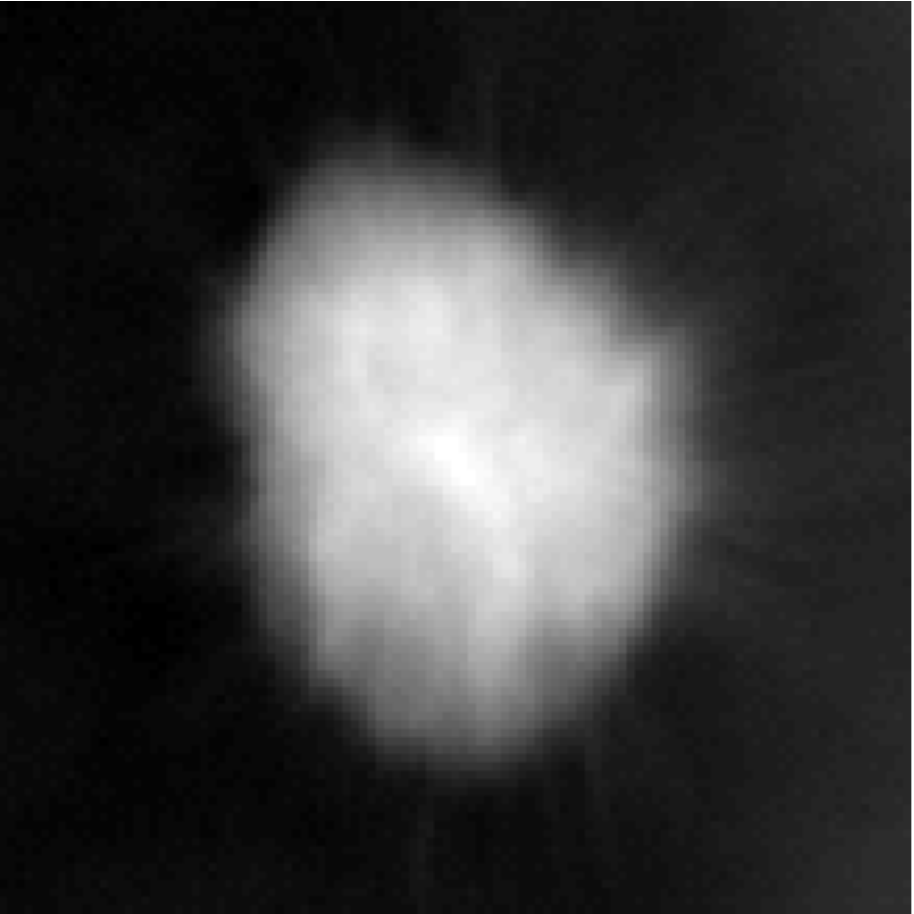}
\caption{}
\end{subfigure}
\begin{subfigure}{0.11\textwidth}
\includegraphics[width=1.0\linewidth]{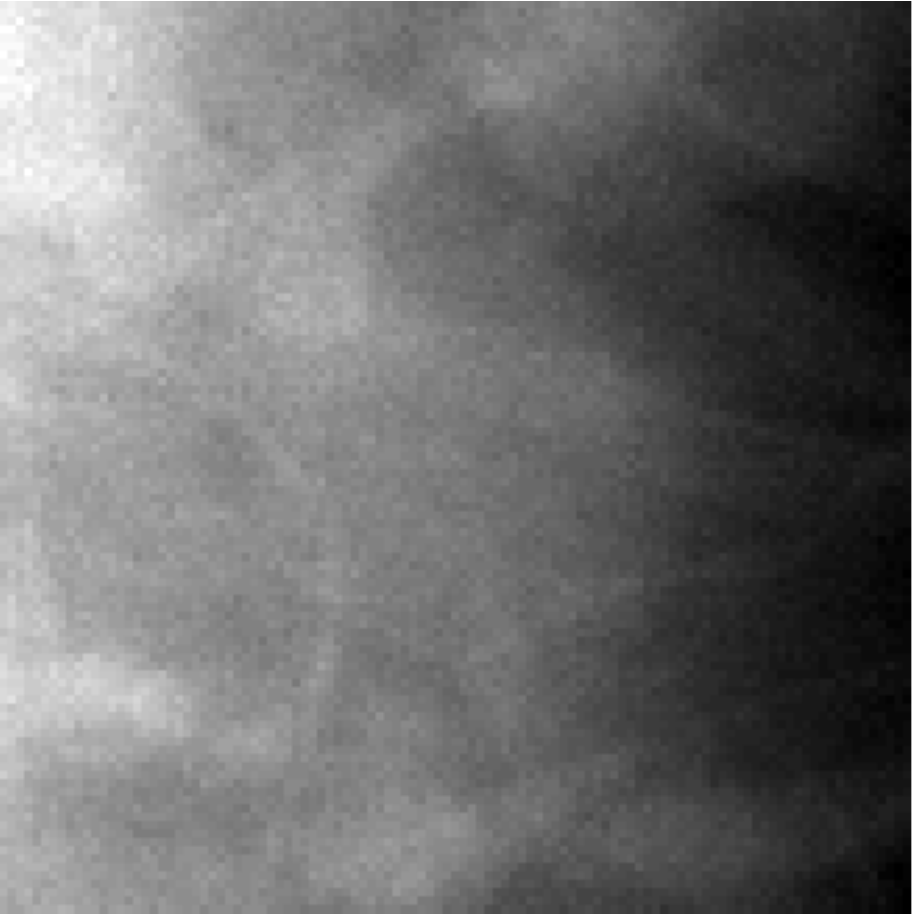}
\caption{}
\end{subfigure}
\begin{subfigure}{0.11\textwidth}
\includegraphics[width=1.0\linewidth]{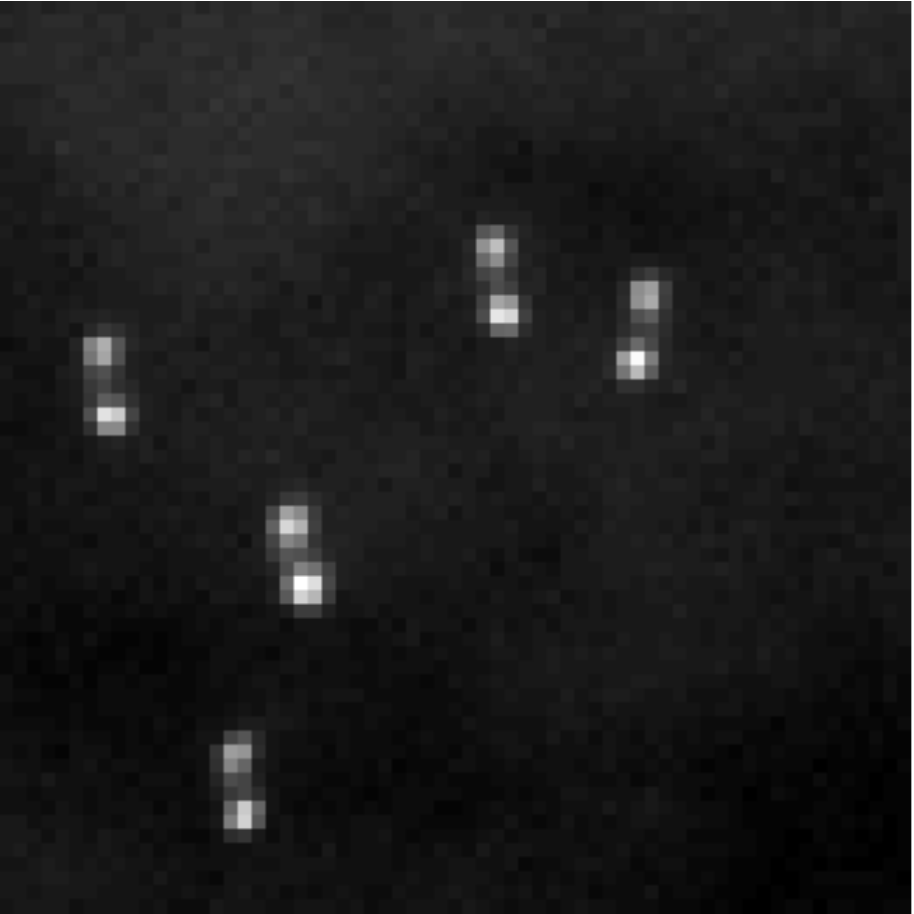}
\caption{}
\end{subfigure}
\begin{subfigure}{0.11\textwidth}
\includegraphics[width=1.0\linewidth]{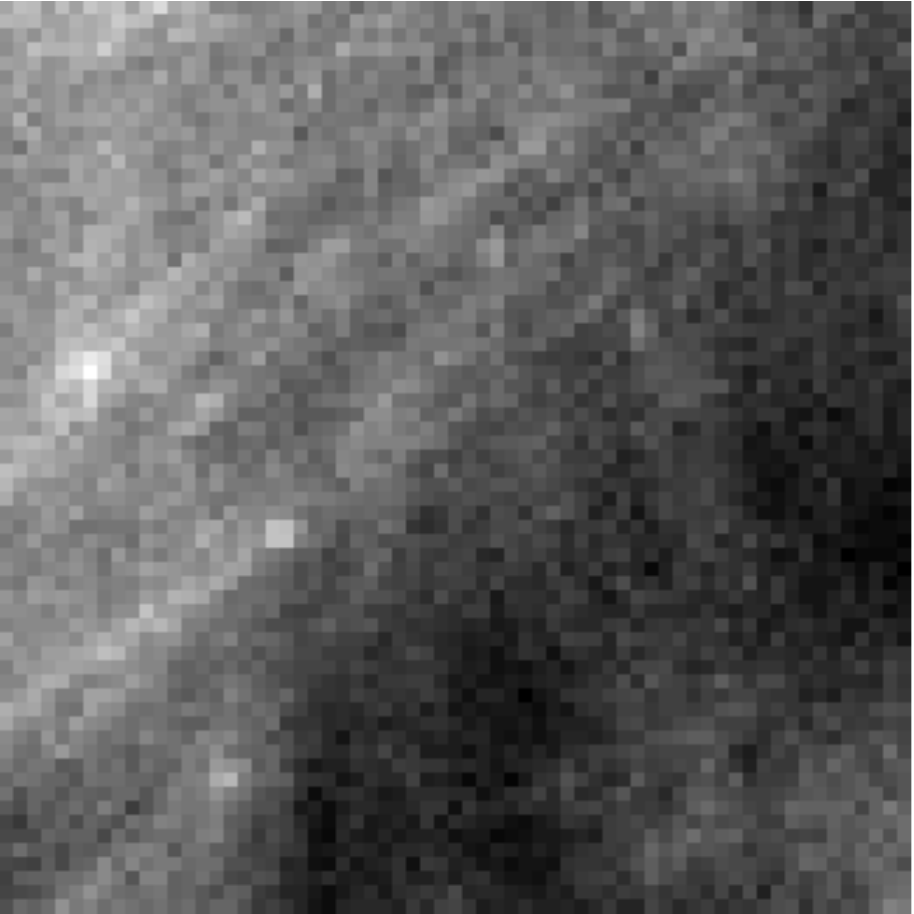}
\caption{}
\end{subfigure}
\caption{Sample estimated signals and images from the VICTRE breast phantom dataset. The grayscale was adjusted in each case to maximize visibility. (a) and (c) contain the mean signal image of the spiculated mass and microcalcification cluster, respectively. (b) and (d) are sample images of those corresponding signals embedded into a fatty breast phantom, which is the easiest detection class in the dataset.}
\label{fig:victre_image}
\end{figure}

For each type of signal, 12500 total images were selected from the dataset to form training, validation, and testing sets of 5000, 625, and 625 paired images, respectively.
The breast types selected maintained the proportions of the VICTRE study~\cite{badano2018victre}.
The signals were estimated by taking the mean of the signal-present images and subtracting the mean of the signal-absent images for the combined training and validation dataset.
Sample images and estimated signals are included in Fig.~\ref{fig:victre_image}.

\subsection{AE Topology}
The considered network topology was a tied-weight AE with no nonlinear or bias terms.
This structure parallels the CHO formulation in Eqn. (\ref{eq:CHO}), as the AE is learning the transformation matrix \(\vec{T}\).
Tied weights were chosen because they couple the encoder and the decoder by enforcing \(\mathbf{W}_1^T = \mathbf{W}_2\), making the encoder a transpose of the decoder.
This formulation prevents loss of information that may solely exist in the decoder since only the encoder is employed as the transformation matrix.
Additionally, tied weight AEs have fewer parameters to train and thus perform better in the limited-data experiments considered~\cite{granstedt2019autoencoder}.



\begin{figure}[t!]
  \centering
  \begin{subfigure}{0.485\textwidth}
  \centering
  \includegraphics[width=1.0\linewidth]{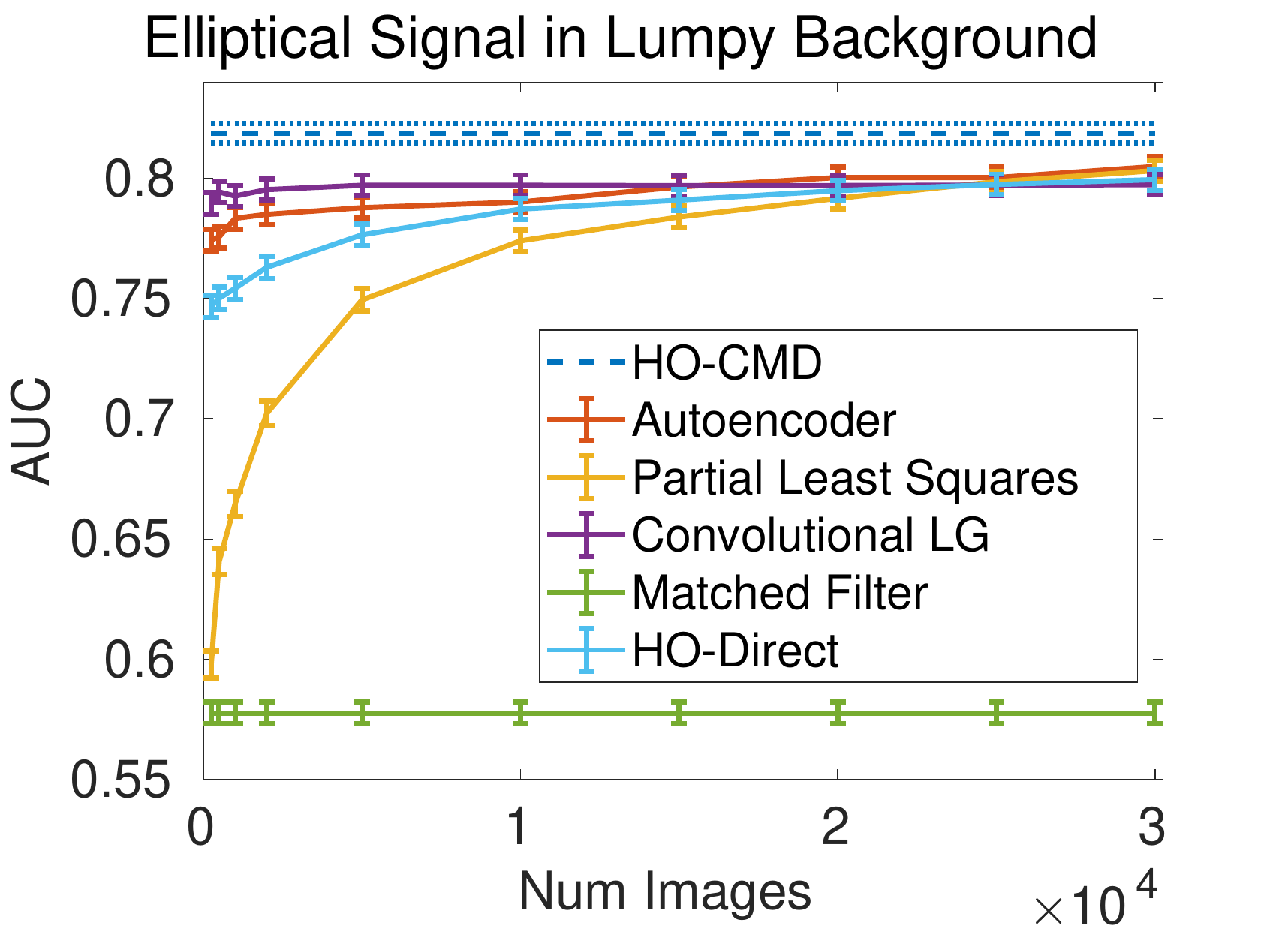}
  \caption{}
  \end{subfigure}
  
  \begin{subfigure}{0.485\textwidth}
  \centering
  \includegraphics[width=1.0\linewidth]{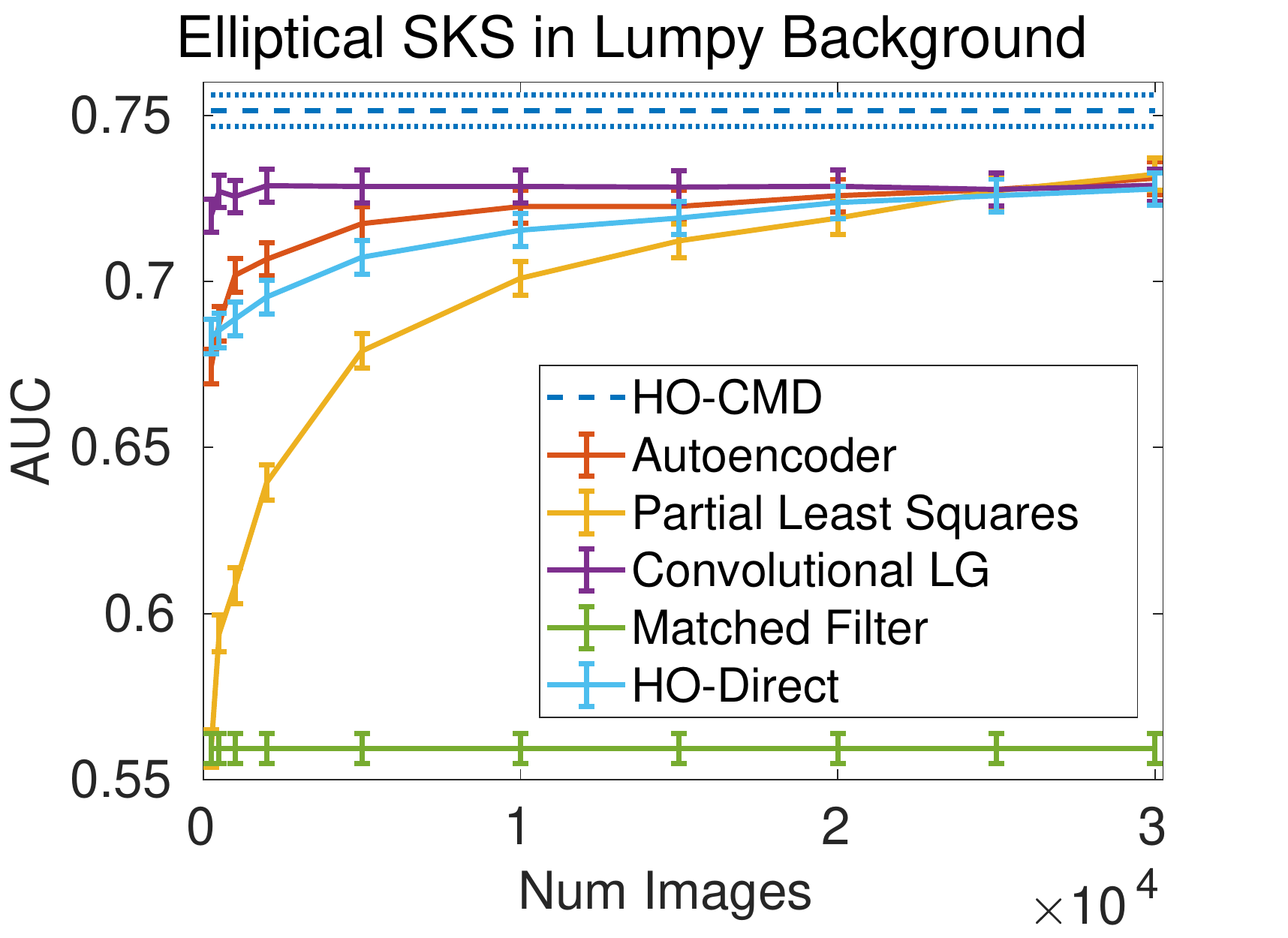}
  \caption{}
  \end{subfigure}
  \caption{Performance of the CHO on varying training dataset sizes for the lumpy background model. (a) contains the results for the location-known elliptical signal while (b) contains the SKS elliptical signal results. The error bars correspond to the standard deviation of the fit AUC values. The HO-CMD is provided as an estimate of the upper bound of the HO, given an infinite amount of images, and is included to benchmark the efficiency of the channels for all methods.}
  \label{fig:lumpy_results}
  \end{figure}
  
  \begin{figure}[t!]
  \centering
  \begin{subfigure}{0.485\textwidth}
  \includegraphics[width=1.0\linewidth]{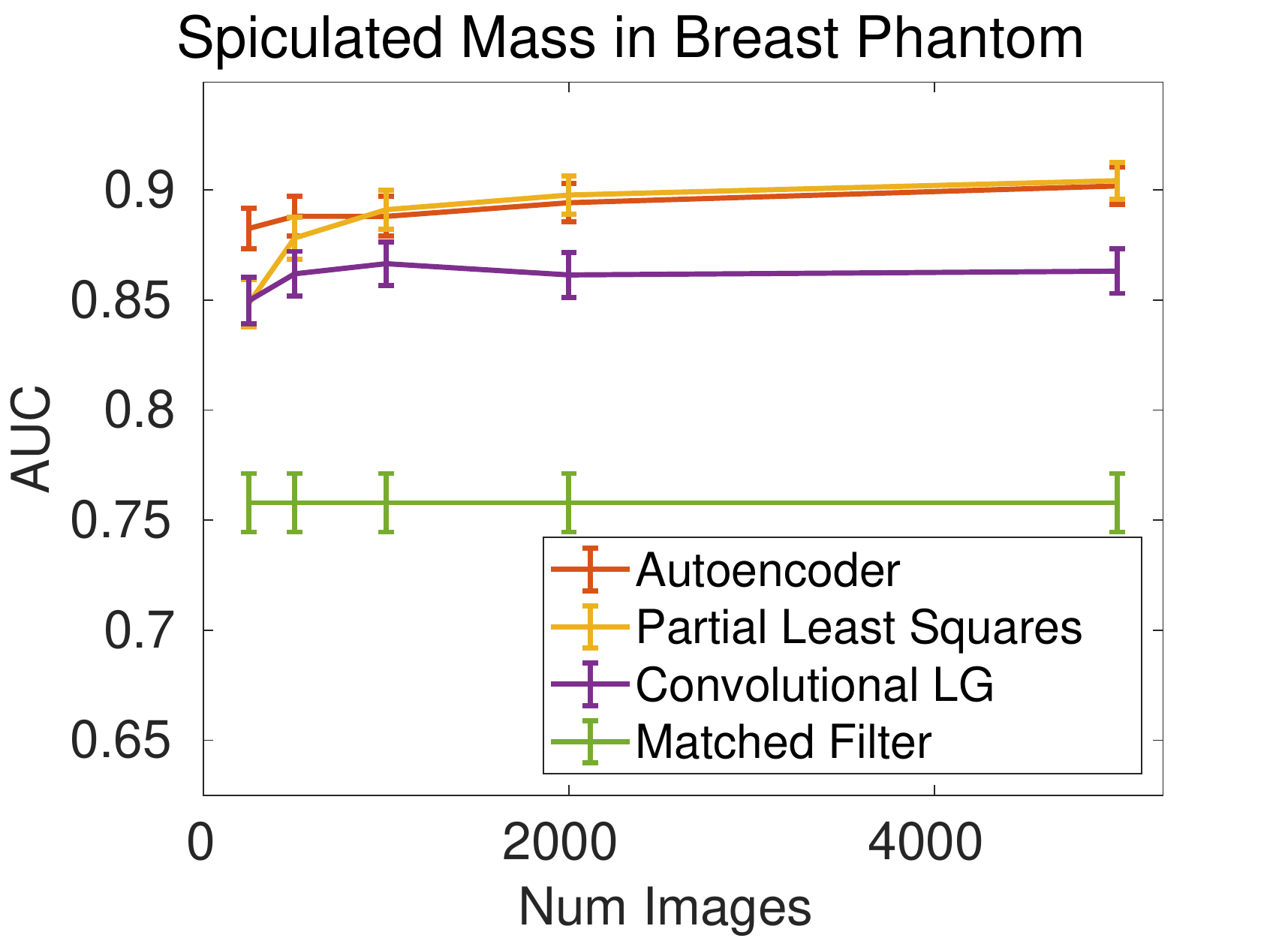}
  \caption{}
  \end{subfigure}
  
  \begin{subfigure}{0.485\textwidth}
  \includegraphics[width=1.0\linewidth]{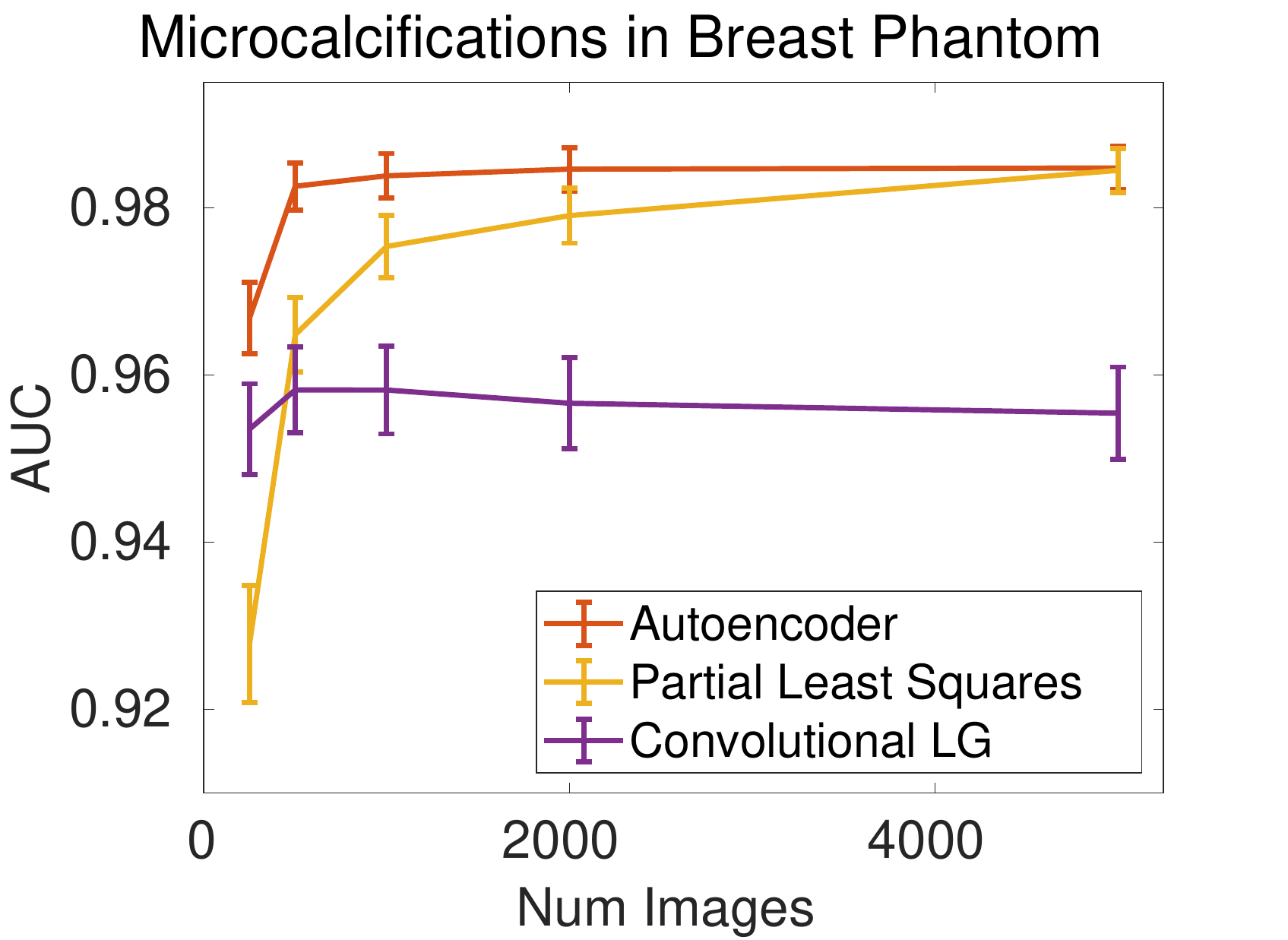}
  \caption{}
  \end{subfigure}
  \caption{Performance of the CHO on varying training dataset sizes for the VICTRE breast phantom model. (a) contains the results for the spiculated mass signal while (b) contains the microcalcification cluster results. The error bars correspond to the standard deviation of the fit AUC values. For both signals, the HO-Direct diverged for most of the dataset sizes due to insufficient data and the more complex background. The analytic HO estimate is not included as the clean images are not available to generate an estimate. The matched filter is also omitted in (b) since it had an AUC of 0.53, significantly less than the channelized methods.}
  \label{fig:clb_results}
  \end{figure}

\subsection{Experimental Parameters}\label{expparam}
\subsubsection{Training Details}
AE-channels were determined by minimizing the modified autoencoder loss function in Eqn. (\ref{eq:AEmod}). %
The models were trained in Tensorflow~\cite{abadi2016tensorflow} using the Adam algorithm~\cite{kingma2014adam}.
The AE weights were initialized using a truncated normal initializer with a standard deviation of 5e-6.
The models were trained for 500 epochs.
Provided the considered dataset contained more than 500 images, pre-training the models on a subset of 500 images for 500 epochs to burn in the network sometimes improved performance.
A mini-batch size of 250 was employed, with an equal number of signal-present and signal-absent images in each mini-batch.
The learning rate was set to 5e-3 for the VICTRE phantom background study and 1e-5 for the lumpy background study.
All networks were trained on a single NVIDIA TITAN X GPU\@.

Several reference methods were implemented to compared against the AE-learned channels, including convolutional LG~\cite{diaz2015derivation}, partial least squares\cite{witten2010pls}, and the matched filter.
The HO-Direct\cite{barrett1993model} was also computed on each subset using Eqn. (\ref{eq:w_HO}).
A grid search on the entire training dataset for each background was used to select the parameters for all methods, with the number of channels capped at 20.
This grid search also implicitly provided multiple random initializations for the AE\@.

\subsubsection{Evaluation}
Each model was on trained across a range of restricted-size subsets of the training data.
The VICTRE case detailed in Sec.~\ref{sec:exp2} contained subsets of size \(K =\) 250, 500, 1000, 2000, and 5000 image pairs.
The larger lumpy background experiments detailed in Sec.\ref{sec:exp1} also considered sets of 10\thinspace000, 15\thinspace000, 20\thinspace000, 25\thinspace000, and 30\thinspace000 image pairs.

The standard train-validate-test scheme~\cite{goodfellow2016deep} was employed to evaluate performance.
The AE and competing methods were given the training data and signal estimate to operate on, with the performance evaluated on the validation data to select the best set of parameters.
Once the parameters were determined for each method, the CHO was numerically determined according to Eqn.~(\ref{eq:HOv}) using the combined training subset and validation dataset to compute \(\vec{K}_\vec{v}\).
The final models were then evaluated on the testing set to obtain the AUC values.

The HO-CMD was also computed for the experiments on lumpy backgrounds to analyze the efficiency of the channels for each method.
The empirical background covariance matrix was calculated using the combined training and validation datasets for a total of 70\thinspace000 noiseless background images. 
This method was unavailable for estimating the HO of the VICTRE experiments as noiseless images were not available.

\section{Results}\label{sec:results}

\begin{figure}
\centering
\includegraphics[width=1.0\linewidth]{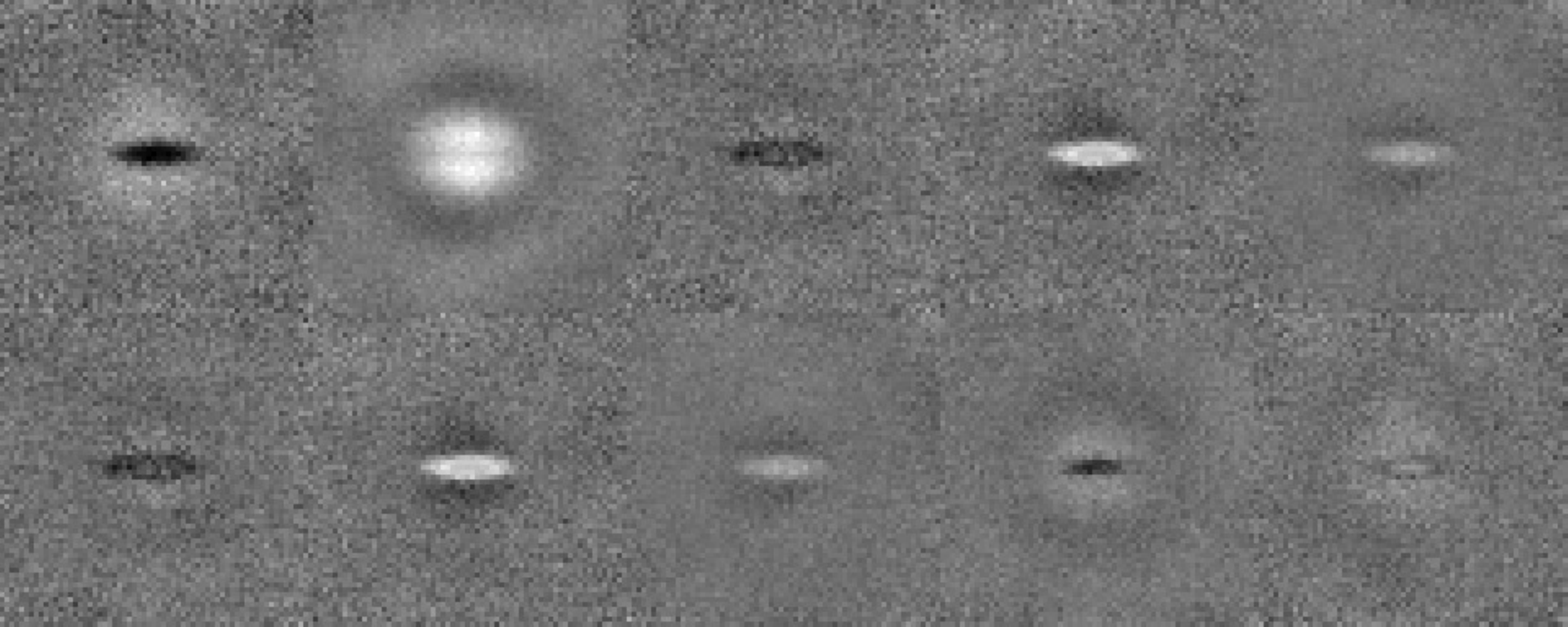}
\caption{Location-known elliptical channels for the 30\thinspace000 paired image dataset on the lumpy background. The grayscale is constant and fixed. Note that some channels have redundant functionality, such as 1 and 6. In this case, removing channel 6 only results in a loss of 0.0005 AUC.}
\label{fig:ae_channels}
\end{figure}

The results for the limited-image tests for the lumpy model and VICTRE breast phantom model are provided in Figs.~\ref{fig:lumpy_results} and~\ref{fig:clb_results}.
The traditional AE was also tested, but failed to exceed 0.55 AUC in all four experiments.
Overall, the proposed method was competitive with the state-of-the-art channelized methods for both the lumpy background and VICTRE phantom background cases.
For the lumpy background cases, the AE-channels performed significantly better than the PLS channels for all but the largest dataset sizes.
In those cases, performance was comparable.
Convolutional LG channel performance was relatively static since the models were tuned at the maximum dataset size and it is not a learning method, but were the best performing channels for the majority of the lumpy dataset sizes considered.
However, both the PLS and AE channels outperformed convolutional LG when sufficient images were available.
The HO-Direct had inferior performance to both the AE and convolutional channels while also requiring significantly more computation to evaluate.
Thus, some channelized methods outperformed the standard method of computing the HO\@.
The HO-CMD serves as an upper bound.

In the VICTRE background case, the AE-learned channels outperform every other tested method for the smaller training subsets.
Given a sufficient amount of data, the AE and PLS channels approach the same AUC and are approximately equivalent. 
This occurred more quickly for the larger spiculated mass signal than the smaller microcalcification clusters.
The HO-Direct also had substandard performance in most cases due to the degeneracy of the covariance matrix in the data-constrained experiments.
In these ill-conditioned cases the test statistic was estimated by solving a linear system, but the resulting low AUC demonstrates the superiority of channelized methods for calculating an observer for this more complicated background.

During the course of the experiments, it was observed that the convolutional LG channels were especially sensitive to the quality of the estimated signal.
When provided with the signal used to generate the data in both the location-known and SKS lumpy experiments, the method outperformed all other competitors.
When fewer images were available and thus there is more noise in the signal image, such as in the VICTRE phantom dataset, the performance degraded significantly.
Although the AE-learned channels attempt to reconstruct the given signal image directly, and thus would seem to be impacted more by noise, the method was more robust to error in the estimated signal than the convolutional LG approach. 
This is likely due to the same innate denoising AEs demonstrate due to the limited embedding dimensionality.

The learned channels for the 30\thinspace000 location-known lumpy image case are included in Fig.~\ref{fig:ae_channels}.
Many of the channels are similar to one another in the features they extract, and can be removed without significant loss of performance.
These extraneous channels likely exist due to the AE training process.
Random initializations generate different starting locations for each channel, which is iteratively optimized by the AE training process.
During this process, the channels are updated to better jointly reconstruct the signal image.
Thus, even if the final model makes inefficient use of its full channel budget, the channels are influenced by their interactions during the training process.
One of the limitations of this approach is its sensitivity to the random initialization, which can result in models of dramatically varying quality even with the same structure.

\section{Discussion and Conclusion}\label{sec:concludes}
This study demonstrated that AEs are capable of learning efficient CHO channels for both location known and certain SKS signal detection tasks.
Data embeddings and observer channels were demonstrated to be fundamentally related, with the task of optimizing a data embedding to preserve signal-specific information equivalent to determining an efficient channel selection for the CHO\@.
Furthermore, the presented method of computing channels is capable of meeting or exceeding the performance of state-of-the-art methods on the investigated tasks.

Channels were learned for the CHO by minimizing the reconstruction loss of an AE\@.
Modification of the AE loss function to focus only on task-specific information involving the signal was found to have a significant benefit over using the traditional AE approach.
Empirical sweeps over the network topology revealed that the AE could efficiently approximate the HO for a wide range of cases utilizing comparable numbers of channels to other approaches.
The proposed method was equivalent to state-of-the-art approaches for the lumpy background and significantly superior on the more complicated VICTRE breast phantom dataset, demonstrating the robustness and versatility of the method.

Performance improvements were especially noticeable for low numbers of training-set images as the AE-learned channels plateaued to higher AUC values sooner than other learning-based methods.
However, the AE-learned channels were sensitive to the random initialization of the weights and frequently learned redundant channels.
The training scheme can likely be further improved with a more robust approach to weight initialization.

Opportunities for future work include expanding the current channels to the IO and extending the formulation to both more sophisticated SKS cases and 3D input images.
The channels should work directly for any standard Markov chain Monte Carlo method for estimating the IO.
Although the current form of the loss function for learning AE-channels requires knowing the signal centroid, it could be generalized by considering convolutional AEs~\cite{lecun2007unsupervised}.
The superior performance of AE-learned channels on smaller datasets and medically realistic phantoms also expands the applicability of the method to real-world cases, and the method should be tested on experimental data to identify remaining challenges in tuning the AE\@.

\section*{Acknowledgment}
This work was supported in part by grants NIH NS102213, NIH EB020604, and NSF DMS1614305.



\ifCLASSOPTIONcaptionsoff
  \newpage
\fi



%
\bibliography{./IEEE_TMI}{}
\bibliographystyle{IEEEtran}


%
%
%
%
%




\end{document}